\documentclass[%
 preprint,
 amsmath,amssymb,
 aps,
 pra,
 12pt
]{revtex4-1}

\usepackage{graphicx}
\usepackage{subfigure}
\usepackage{bm}
\usepackage[mathlines]{lineno}
\usepackage{color}
\usepackage{psfrag}
\usepackage {CJK}
\usepackage{latexsym}
\usepackage{amssymb}
\usepackage{indentfirst}
\usepackage[letterpaper]{geometry}
\usepackage{euscript}
\usepackage{fullpage}
\usepackage{mathrsfs}
\usepackage{amsfonts}

\usepackage{braket}

\usepackage{verbatim}
\usepackage{makecell}

\graphicspath{{figure/}}

\newcommand{\EQ}{\begin{equation}}
\newcommand{\EN}{\end{equation}}
\newcommand{\EQA}{\begin{eqnarray}}
\newcommand{\ENA}{\end{eqnarray}}

\newcommand{\Real}{\operatorname{Re}}

\newcommand{\rnorm}[1]{\left\langle#1\right\rangle_{\mathbb{R}}}

\newcommand{\figref}[1]{Fig.~\ref{#1}}
\renewcommand{\eqref}[1]{Eq.~(\ref{#1})}

\begin{document}

\begin{CJK*}{UTF8}{gbsn}

\title{Quantum state preparation for a velocity field based on the spherical Clebsch wave function}

\author{Hao Su} 
 \affiliation{State Key Laboratory for Turbulence and Complex Systems, College of Engineering, Peking University, Beijing 100871, China}
\author{Shiying Xiong} 
\email{shiying.xiong@zju.edu.cn}
 \affiliation{Department of Engineering Mechanics, School of Aeronautics and Astronautics, Zhejiang University, Hangzhou 310027, China}
\author{Yue Yang} 
 \email{yyg@pku.edu.cn}
 \affiliation{State Key Laboratory for Turbulence and Complex Systems, College of Engineering, Peking University, Beijing 100871, China}%
 \affiliation{HEDPS-CAPT, Peking University, Beijing 100871, China}

\date{\today}

\begin{abstract}
We propose a method for preparing the quantum state for a given velocity field, e.g., in fluid dynamics, via the spherical Clebsch wave function (SCWF). 
Using the pointwise normalization constraint for the SCWF, we develop a variational ansatz comprising parameterized controlled rotation gates. Employing the variational quantum algorithm, we iteratively optimize the circuit parameters to transform the target velocity field into the SCWF and its corresponding discrete quantum state, enabling subsequent quantum simulation of fluid dynamics. 
Validations for one- and two-dimensional flow fields confirm the accuracy and robustness of our method, emphasizing its effectiveness in handling multiscale and multidimensional velocity fields. Our method is able to capture critical flow features like sources, sinks, and saddle points. Furthermore, it enables the generation of SCWFs for various vector fields, which can then be applied in quantum simulations through SCWF evolution. 
\end{abstract}

\maketitle
\end{CJK*}

\section{Introduction}

Quantum computing exploits the principles of quantum mechanics, like superposition and entanglement, to encode information and perform data operations using qubits \cite{Steane1998, Nielsen2010, Horowitz2019}. Its significant potential for speeding up specific computational problems has attracted immense interest among researchers across diverse fields \cite{Cao2019, Wittek2014, Mavroeidis2018, Schuld2015, Jin2023}. 
In particular, in specific scenarios, quantum computing of fluid dynamics (QCFD) could be more efficient than the methods in classical computational fluid dynamics (CFD)~\cite{Gaitan2020}. 

Similar to pre-processing in CFD,  the initial state preparation in QCFD converts classical information into quantum information encoded with qubits.  
This step is crucial as it directly affects the performance and reliability of QCFD.

The initial state preparation in QCFD shares common challenges with other quantum computing problems, as previous research has suggested that accurately preparing a quantum state usually needs a large number of quantum gates, thereby making the algorithm impractical. 
Specifically,  Shende \textit{et al.}~\cite{Shende2004} estimated that for an $n$-qubit system, an exponentially large number of elementary quantum gates, $O(4^n)$, are required to ensure the preparation of a given unitary matrix. Subsequent studies~\cite{Shende2006, Drury2008, Khammassi2021} have proposed various methods for decomposing unitary matrices using elementary quantum gates at an exponential scale.
However, this remains impractical for implementation on contemporary noisy intermediate-scale quantum (NISQ) hardware~\cite{Preskill2018, Bharti2022}. Further research indicates that designing circuits tailored to specific problems 
can reduce the number of quantum gates~\cite{Khatri2019}. 

In particular, the initial state preparation in QCFD faces unique difficulties in encoding classical fields as quantum states. Classical fluid dynamics describes fluid motion using a continuously distributed velocity or vorticity field~\cite{Batchelor1967}. Consequently, preparing initial states in QCFD requires converting these vector fields into quantum states through a ``quantization'' process~\cite{nair2016elements}. 
One approach to address this problem simplifies nonlinear fluid dynamics into a series of linear differential equations~\cite{Gaitan2020, Ray2022, Pfeffer2022}, which are then solved using quantum linear solvers~\cite{Harrow2009}. However, this linearization struggles to accurately describe turbulence and transition phenomena commonly present in real fluid flows. 
To better capture the evolution of classical fluid flow with viscous vortices, Meng \textit{et al.}~\cite{Meng2023, Meng2024quantum} introduced the Hamiltonian simulation for QCFD. This method employs a generalized Madelung transformation to represent fluid velocity via a spherical Clebsch wave function (SCWF)~\cite{Chern2016,Chern2017,Yang2021clebsch,Xiong2022Clebsch,Tao2021}. 
However, obtaining the SCWF from a given velocity field remains an open problem.

We aim to devise a general method for preparing quantum states in QCFD, based on the SCWF and the variational quantum algorithm (VQA)~\cite{Benedetti2019, Jaksch2023}. The prepared initial quantum state serves as the initial state for subsequent state evolution. 
We formulate the SCWF construction as an optimization problem subject to a normalization constraint, with a loss function linking the SCWF and the velocity field.
In the implementation, we devise a tailored quantum circuit with SCWF normalization, utilizing Hadamard gates and parametrized controlled rotation gates. 
%
Our method is end-to-end, i.e., directly constructing quantum states for Hamiltonian simulation from a given velocity field. By utilizing VQA-optimized quantum circuits, it enables a linear growth of the number of quantum gates with the number of qubits, avoiding the exponential gate count in the existing general state preparation algorithms. 
%

The outline of this paper is as follows. 
Section \ref{sec:basic_knowledge} introduces the SCWF and formulates the optimization problem for constructing the SCWF. 
Section \ref{sec:quantum} details the specific settings of the VQA for constructing the SCWF and its corresponding quantum circuit, along with a discussion on computational complexity.
Section \ref{sec:result} assesses our algorithm using 1D Fourier series and 2D flow fields.  
Conclusions and future works are discussed in Section \ref{sec:conclusion}.

\section{Constructing SCWF with VQA}\label{sec:basic_knowledge}

\subsection{Construction of SCWF}\label{sec:spherical_Clebsch}
The SCWF $\bm{\psi}(\bm{x},t)=[\psi_1(\bm{x},t),\psi_2(\bm{x},t)]^T$ is a quantum representation of a velocity field, where $\psi_i(\bm{x},t)=a_i(\bm{x},t)+\textrm{i} b_i(\bm{x},t)$ (for $i =1,2$) denote two complex-valued functions, with real-valued functions $a_i(\bm{x},t)$ and $b_i(\bm{x},t)$. 
The SCWF is required to be normalized, i.e., it satisfies
\begin{equation}\label{eq:wavefunction_normalization}
|\bm{\psi}|^2=\rnorm{\bm{\psi},\bm{\psi}}=1.
\end{equation}
Here, the inner product $\rnorm{\cdot ,\cdot}$ is defined as  $\rnorm{\bm{\psi},\bm{\phi}}\equiv\Real(\overline{\psi}_1\phi_1+\overline{\psi}_2\phi_2)$, where $\overline{\psi}_i$ denotes the complex conjugate of $\psi_i$. 
The relation between the velocity and the SCWF~\cite{Chern2017thesis} is described by
\begin{equation}\label{eq:psi_to_u}
\bm{u}_{\psi}\equiv \hbar\rnorm{\bm \nabla\bm{\psi},\textrm{i}\bm{\psi}} = \hbar(a_1\bm \nabla b_1-b_1\bm \nabla a_1+a_2\bm \nabla b_2-b_2\bm \nabla a_2),
\end{equation}
where $\hbar$ is a constant, and its physical interpretation can be found in Ref.~\cite{Chern2016}. 

Through the Hopf fibration~\cite{Hopf1931}, the SCWF can be transformed into the spin vector $\bm{s}=(s_1,s_2,s_3)$ on the Bloch sphere \cite{Bloch1946}, with 
\begin{equation}\label{eq:Hopf_fibration}
s_1=a_1^2+b_1^2-a_2^2-b_2^2,~~
s_2=2(a_2b_1-a_1b_2),~~
s_3=2(a_1a_2+b_1b_2).
\end{equation}
Each $s_i$ is a vortex-surface field \cite{Yang2010b,Yang2011b} for the vorticity $\bm{\omega}_{\psi}\equiv\bm \nabla\times\bm{u}_{\psi}$, with the constraint $\bm{\omega}_{\psi}\cdot\bm \nabla s_i=0$, $i=1, 2, 3$. 
This mapping shows that each vortex line in Euclidean space is associated with a point on the Bloch sphere, facilitating the visualization \cite{Chern2017} and simulation \cite{Yang2021clebsch,Xiong2022Clebsch} of various vortical flows. 

We aim to develop a quantum preparation algorithm for calculating the SCWF $\bm \psi$ that satisfies \eqref{eq:psi_to_u} for a given velocity field. Nevertheless, it is usually not feasible to solve \eqref{eq:psi_to_u} directly, 
so we reformulate the inverse problem of \eqref{eq:psi_to_u} as an optimization problem
\begin{equation}\label{eq:inverse_problem}
\bm \psi_{u} = \mathop{\arg\min}\limits_{\|\bm{\psi}\|=1}\frac{1}{\hbar^2}\|\bm{u}_{\psi}-\bm{u}\|_{\epsilon}^2,
\end{equation}
where 
the norm $\|\cdot\|_{\epsilon}$ incorporates a regularization term \cite{Chern2016} as 
\begin{equation}
\|\bm{u}_{\psi}-\bm{u}\|_{\epsilon}^2 = \|\bm{u}_{\psi}-\bm{u}\|^2 + \epsilon^2 \left\|\hbar\bm \nabla \bm \psi - \textrm{i}\bm u_{\psi}\bm \psi \right\|^2.
\label{eq:LGnorm}
\end{equation}
The regularizer $\epsilon^2 \left\|\hbar\bm \nabla \bm \psi - \textrm{i}\bm u_{\psi}\bm \psi \right\|^2$ enhances convexity of the optimization to obtain a unique solution. 
We dynamically adjust $\epsilon$ to balance the solution accuracy and problem regularization. 
Additionally, both the wave function and the velocity have periodic boundary conditions in the present study. 
Further clarification on the optimization problem in \eqref{eq:inverse_problem} and its classical algorithms are provided in Ref.~\cite{Chern2017}.

\subsection{VQA for preparation circuit}\label{sec:VQA_intro}

\begin{figure}
    \centering
    \includegraphics[width=0.9\textwidth]{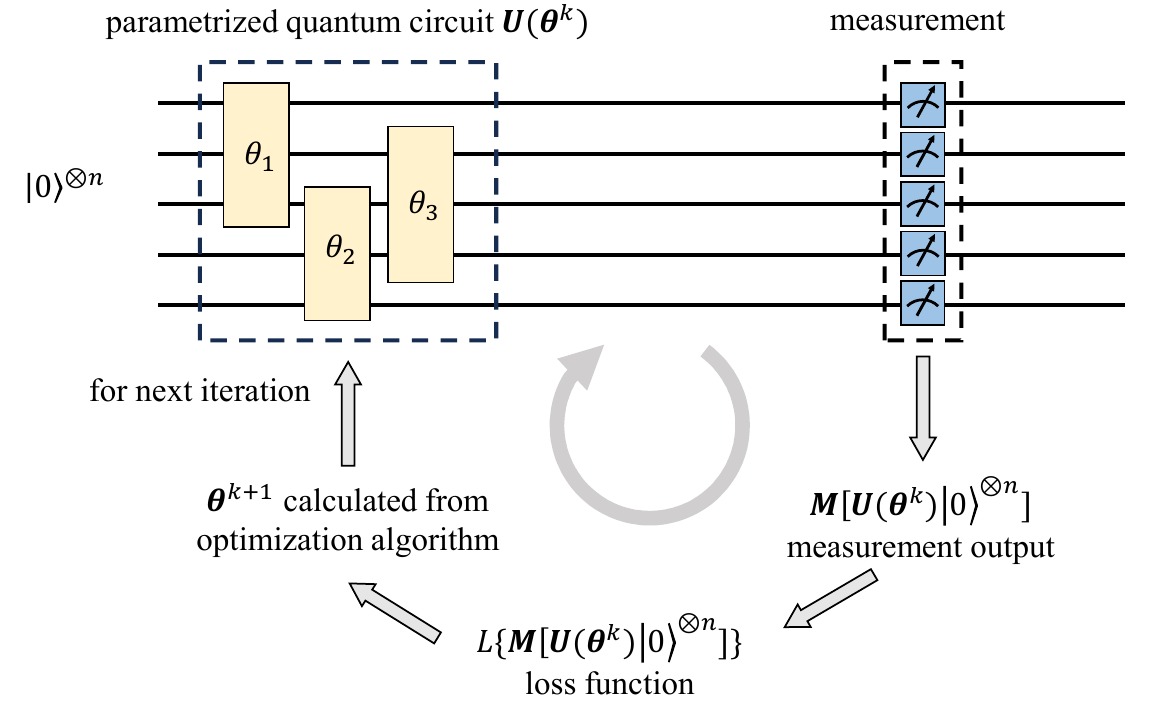}
    \caption{The VQA diagram illustrates the iteration from the default initial state $\ket{0}^{\otimes n}$ to an output quantum state $\ket{\bm{\psi}}$ via tuning a parameterized quantum circuit $\bm U(\bm \theta)$, expressed as $\ket{\bm{\psi}}=\bm U(\bm \theta)\ket{0}^{\otimes n}$. The circuit parameters $\bm \theta$ are tuned using the AdamW optimization algorithm \cite{Loshchilov2019}, optimizing with respect to a loss function $L(\cdot)$. Here, $\bm{\theta}^k$ denotes the parameter values at the $k$-th iteration.}
    \label{fig:VQA}
\end{figure}

We discretize both the velocity and SCWF on a uniform grid. 
Within the VQA framework, we iteratively update parameters within quantum circuits to obtain the target preparation circuit \cite{Benedetti2019}.
As sketched in Fig.~\ref{fig:VQA}, the VQA procedure is similar to a machine learning problem. The circuit incorporates parametrized quantum operations $\bm U(\bm \theta)$, which are influenced by multiple independent parameters represented as a vector $\bm \theta$. This quantum circuit can be conceptualized as a unitary matrix, transforming a given state into a desired initial state.

A loss function $L(\cdot)$ is formulated based on the quantum state output, attaining its minimum value when the desired output state is achieved. The optimization is then formulated by varying parameters $\bm \theta$ to optimal ones
\begin{equation}
\bm \theta^{\textrm{opt}}=\mathop{\arg\min}\limits_{\bm \theta} L \{\bm M[\bm U(\bm \theta) \ket{0}^{\otimes n}]\}
\end{equation}
to minimize the loss function, where $n$ denotes the number of qubits, $\ket{0}^{\otimes n}$ denotes the default initial state, and $\bm M$ represents the quantum state measurement. 
Note that the output of the quantum circuit was obtained and the circuit parameters were optimized on a classical computer.

We encode the SCWF with $n$ qubits using a parameterized quantum circuit
\begin{equation}
\ket{\bm \psi} = \bm U(\bm \theta)\ket{0}^{\otimes n}.
\end{equation}
The components of $\ket{\bm \psi}$ store the SCWF at various grid points. 
We define the loss function
\begin{equation}
\label{eq:loss2}
L =\frac{1}{\hbar^2}\| \bm u_{\psi}-\bm u_t\|_{\epsilon}^2
\end{equation}
using a target velocity $\bm{u}_t$. 
This loss function corresponds to the optimization problem described in \eqref{eq:inverse_problem}, where $\bm \psi$ is obtained from the quantum circuit and can be converted into $\bm u_{\psi}$ using \eqref{eq:psi_to_u}. 
We then iteratively adjust $\bm \theta$ using the AdamW optimization algorithm~\cite{Loshchilov2019} to minimize \eqref{eq:loss2} and solve \eqref{eq:inverse_problem}.

\section{Implementation of the VQA}\label{sec:quantum}

\subsection{Quantum encoding of SCWFs}\label{sec:encoding}

We consider a $d$-dimensional domain $[0,L]^d$ with length $L=2\pi$, discretized by $N^d$ uniform grid points. SCWFs are defined at grid points $(i_1\Delta,i_2\Delta,\cdots,i_d\Delta)$, where $\Delta=L/N$ denotes the grid spacing and $i_k$ ranges from $0$ to $N-1$ with $k=1,2,\cdots,d$. These SCWFs,  representing a discretized vector field, are encoded with $n$ qubits. 
Specifically, one qubit is used to encode two SCWF components, and the other $n-1$ ones encode SCWF positions, satisfying the condition $2^{n-1}=N^d$. 

Following the encoding method of Meng and Yang \cite{Meng2023}, we discretize a SCWF as
\begin{equation}\label{eq:coding}
\ket{\bm{\psi}}=\dfrac{1}{\mathcal{N}}\sum_{j_{n-2}=0}^1\sum_{j_{n-3}=0}^1\cdots\sum_{j_0=0}^1[\psi_1(\bm x_j)\ket{0}+\psi_2(\bm x_j)\ket{1}]\otimes\ket{j_{n-2}j_{n-3}\cdots j_0}.
\end{equation}
Here, the initial $n-1$ qubits $\ket{j_{n-2}j_{n-3}\cdots j_0}$ encode the binary representation of the index $j$ corresponding to the spatial coordinate $\bm x_j$. 
For each grid point at $\bm x_j$, the two components $\psi_1$ and $\psi_2$ are distinguished by the $n$-th qubit as $\psi_1(\bm x_j)\ket{0}+\psi_2(\bm x_j)\ket{1}$. 
The normalization coefficient $1/\mathcal{N}$ with $\mathcal{N}=\sqrt{2^{n-1}}$ ensures that the entire quantum state has a unity norm. 

\subsection{Designing quantum circuit for state preparation}
\begin{figure}
    \centering
    \includegraphics[width=1.\linewidth]{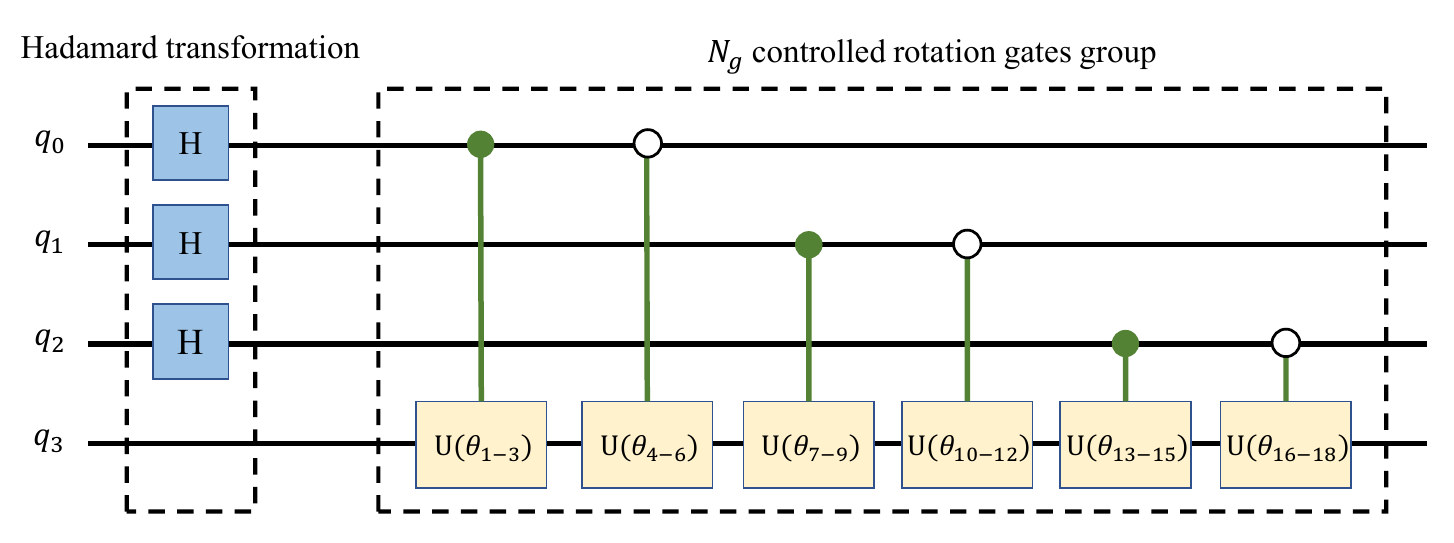}
    \caption{Schematic of a parameterized quantum circuit with enforced pointwise normalization in \eqref{eq:wavefunction_normalization}. The circuit employs Hadamard transformation and controlled rotation gates. The Hadamard transformations ensure the SCWF normalization across $n-1$ qubits. To maintain the normalization, no further operations are conducted on these qubits after the Hadamard transformation. Instead, controlled rotation gates are applied to the $n$-th qubit, so that the former $n-1$ qubits serve as control qubits. The gate's action on the target qubit is controlled by $\ket{1}$ (solid circles) and $\ket{0}$ (open circles).
    }
    \label{fig:circuit_structure}
\end{figure}

In general, quantum circuits are not subject to the pointwise normalization condition in \eqref{eq:wavefunction_normalization} for the two-component wave function. 
Thus, we devise a 
specialized quantum circuit to ensure that any variation of $\bm \theta$ fulfills the pointwise normalization condition in the resulting quantum state.
The quantum circuit is illustrated using a four-qubit system in Fig.~\ref{fig:circuit_structure}.  
This circuit can be extended for more qubits with a similar structure. 

The quantum circuit comprises a parameter-free Hadamard transformation and controlled rotation gates with adjustable parameters. First, we perform Hadamard transformations on $n-1$ qubits. 
This process guarantees that the default input is transformed into a quantum state with pointwise normalization condition specified in \eqref{eq:wavefunction_normalization}~\cite{Nielsen2010}.
To maintain the normalization condition of the SCWF, we do not conduct further operations on the $n-1$ qubits after the Hadamard transformation. 
Instead, controlled rotation gates are employed on the $n$-th qubit in the remaining part of the preparation circuit.  
In this way, one of the former $n-1$ qubits serves as control qubits.

As shown in Fig.~\ref{fig:circuit_structure}, we employ two types of control qubits to manipulate the rotation qubits, identified by solid and open circles, respectively.
The gate applies the rotation matrix to the target qubit if the control qubit reads $\ket{1}$ for the solid circle, and $\ket{0}$ for the open circle.

The controlled rotation gates, following the section consisting of Hadamard transformations, are general one-qubit controlled-unitary gates. 
All of them are $2 \times 2$ complex unitary matrices, each comprising three independent parameters.
The arrangement of the controlled rotation gates is flexible. Here we organize these gates into multiple groups. 
Each group is generated by enumerating all possible control qubits, resulting in $2(n-1)$ distinct gates. This group is replicated $N_g$ times and added to the circuit. 
For example, the circuit with $n=4$ and $N_g=2$ has a total of $2(n-1)N_g=12$ gates in Fig.~\ref{fig:circuit_structure}.

\subsection{Training setup and computational complexity}\label{sec:loss_function}

%
We perform numerical optimization of circuit parameters using VQA on a classical computer. This procedure involves executing circuits and computing gradients of the loss function $L$ in \eqref{eq:loss2} with respect to parameters $\bm{\theta}$.
We conducted our method using the PyTorch programming framework, including PyTorch's automatic differentiation for gradient computation, and the AdamW optimizer for parameter optimization. This implementation is recommended as effective for VQA training~\cite{Lockwood2022}.

For calculating $L$ defined in \eqref{eq:loss2}, the initial learning rate is set at 0.03 and decreases to 0.015 throughout the optimization process. The regularization parameter $\epsilon$ begins at 1 and is reduced by 30\% every 1000 iterations. 
Note that as the numbers of qubits and parameters increase, the barren plateau phenomenon, characterized by gradient disappearance during optimization, becomes prevalent~\cite{Mcclean2018, Martin2023, Liu2022}. Our training methodology partially addresses this challenge by incorporating a regularization term and imposing the pointwise normalization constraint on the circuit output.

As illustrated in Fig.~\ref{fig:circuit_structure}, for $n$ qubits with $N_g$ groups, the number of quantum gates in our circuit is 
\begin{equation}\label{eq:gate_count}
    N_q=2(n-1)N_g\sim O(nN_g). 
\end{equation}
This is linear to $n$ for a constant $N_g$, and it can be much smaller than the exponential estimate of $N_q = O(4^n)$ for a general initial state~\cite{Shende2004}.

We examine the time complexity for our initial preparation algorithm. The grid, containing \( N^d \) points, corresponds to $n$ qubits with the relation \( 2^{n-1} = N^d \). 
During each iteration, evaluating the loss function involves computing the gradient term \( \bm\nabla \bm{\psi} \) in \eqref{eq:LGnorm}. 
The gradient is calculated using the central difference across all \( N^d \) grid points, requiring \( O(N^d) \) operations. 
Therefore, the calculation of $L$ takes \( O(N_qN^d)\sim O((n-1)2^{n}N_g) \) operations per step. 

For each optimization step, the gradient backpropagation is performed for every controlled rotation gate. Each gate applies a transformation on half of the total $2^n$ quantum states, resulting in a complexity of \( O(2^{n-1}\cdot 2(n-1)N_g )\sim O(2^n(n-1)N_g)\) for a single optimization step.
For the entire training procedure with $N_i$ iteration steps, the time complexity is
\begin{equation}\label{eq:complexity}
    T(n)\sim O(2^n[(n-1)+(n-1)]N_gN_i)\sim O(n2^{n+1}N_gN_i),
\end{equation}
which is exponential to $n$. 

Furthermore, the definition of $L$ encompasses all quantum state components of the circuit output, suggesting that computing and optimizing $L$ necessitate measurements of all quantum state components, with complexity increasing exponentially with $n$, thereby presenting significant technical challenges~\cite{Nielsen2010}.

In summary, our algorithm exhibits linear scalability in quantum gate count with respect to the number of qubits, indicating the potential for accelerated execution of trained quantum circuits on quantum hardware. However, the training process can be time-consuming, because the present loss function has to be computed by measuring and computing all quantum state components. 
It is anticipated that improving the loss function will enhance training efficiency.

\section{Results}\label{sec:result}
 
We demonstrate the feasibility and scalability of our method through 1D ($d=1$) and 2D ($d=2$) case studies. 
Detailed parameters and relative errors for all cases are listed in Tab.~\ref{table:config}. 
Here, the average relative error is   
\begin{equation}\label{eq:relative_error}
\varepsilon(\bm{u}_{\psi},\bm{u}_t)\equiv\dfrac{\mathrm{Avg}(|\bm u_{\psi}-\bm u_t|)}{\mathrm{Avg}(|\bm u_t|)},
\end{equation}
where $\mathrm{Avg}(\cdot)$ gives mean value over all grid points. This error measures the discrepancy between the SCWF-based velocity $\bm{u}_{\psi}$ and the target velocity $\bm{u}_t$. Furthermore, a sensitivity analysis with different parameter values is provided in Appendix \ref{sec:parameters}.

\begin{table}
    \begin{center}
    \caption{Parameters and errors in our testing cases, where $\bm{u}_t$ (bold for a vector in multiple dimensions, non-bold for one dimension) represents the target velocity, $d$ denotes the number of dimensions, $n$ the number of qubits, $N_g$ the number of groups of controlled rotation gates, $N$ the number of grid points along each dimension, $\Omega$ the computational domain, $\hbar$ a factor associated with quantifying the vorticity field, and the relative error $\varepsilon(\bm{u}_{\psi}, \bm{u}_{t})$ is defined in \eqref{eq:relative_error}.}
    \label{table:config}
        \begin{tabular}{ccccccccc}
            \hline
            Case\quad\qquad&$\bm u_t$                     \quad\qquad&$d$\qquad\qquad&$n$\qquad\qquad&$N_g$\qquad\qquad&$N$\qquad\qquad&$ \Omega$   \qquad\qquad&$\hbar$\qquad\qquad&$\varepsilon(\bm{u}_{\psi},\bm{u}_{t})$\\\hline
            1   \quad\qquad&$\sin{x}$                     \quad\qquad&1  \qquad\qquad&6  \qquad\qquad&2    \qquad\qquad&32 \qquad\qquad&$[0,2\pi]$  \qquad\qquad&$1$    \qquad\qquad&3.83$\%$\\
            2   \quad\qquad&$\sin{x}+\cos{2x}+\sin{3x}$   \quad\qquad&1  \qquad\qquad&6  \qquad\qquad&2    \qquad\qquad&32 \qquad\qquad&$[0,2\pi]$  \qquad\qquad&$1$    \qquad\qquad&8.58$\%$\\
            3   \quad\qquad&$[\cos x\sin y, \sin x\cos y]$\quad\qquad&2  \qquad\qquad&11 \qquad\qquad&5    \qquad\qquad&32 \qquad\qquad&$[0,2\pi]^2$\qquad\qquad&$1$    \qquad\qquad&4.50$\%$\\
            \hline
        \end{tabular}
    \end{center}

\end{table}

\subsection{1D Fourier series}
\begin{figure}
    \centering
    \subfigure{
        \centering
        \includegraphics[width=0.31\linewidth]{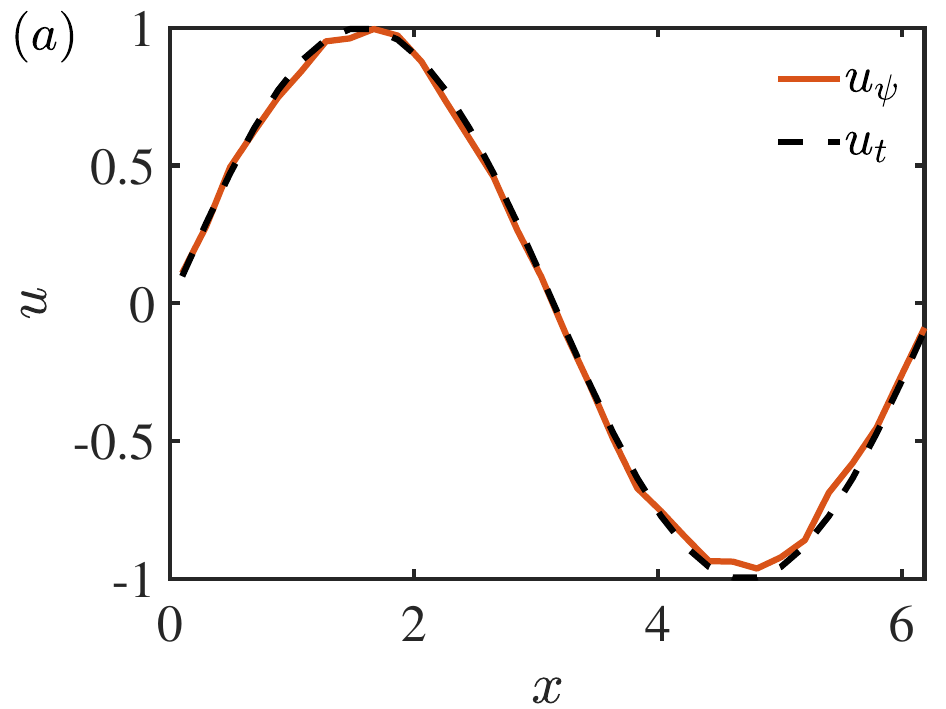}
        \label{fig:sin_qis_u}
    }
    \subfigure{
        \centering
        \includegraphics[width=0.31\linewidth]{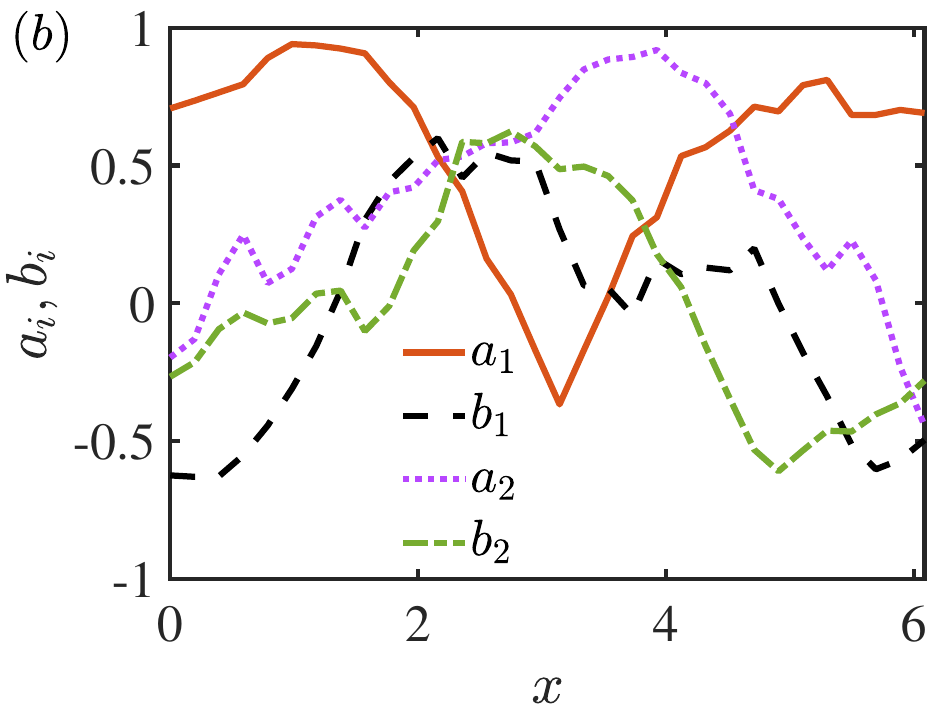}
        \label{fig:sin_qis_abcd}
    }
    \subfigure{
        \centering
        \includegraphics[width=0.31\linewidth]{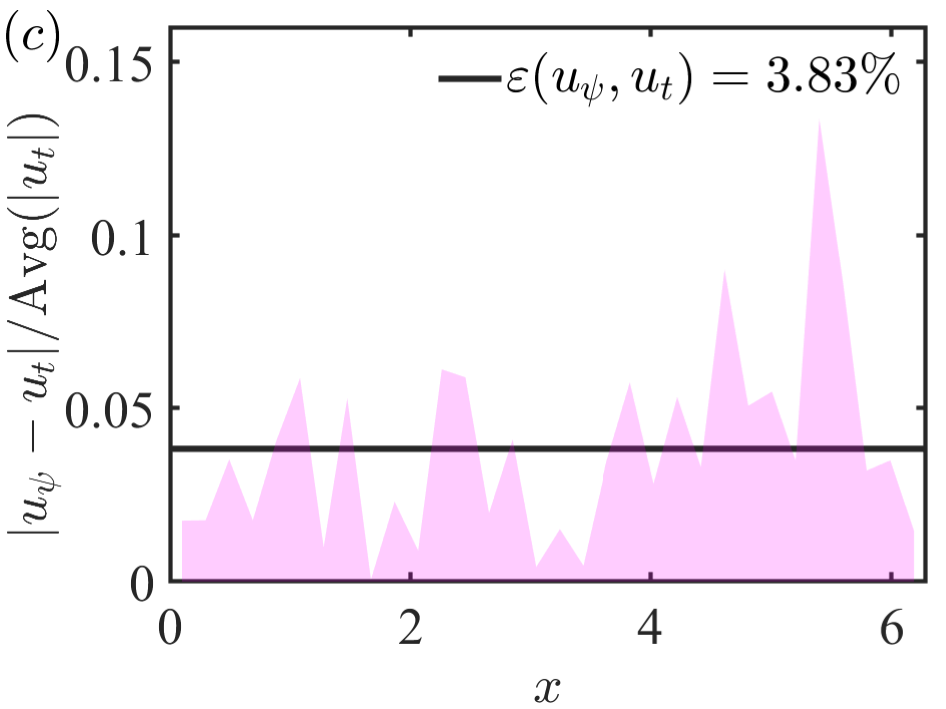}
        \label{fig:sin_qis_dev}
    }
    \caption{Results for generating the SCWF for $u_t=\sin x$. 
    (a) Comparison between the target velocity $u_t$ (black dashed line) and the quantum optimization result $u_{\psi}$ (red solid line). 
    (b) Four real components of the SCWF $\bm{\psi}=[a_1+b_1{\rm i},a_2+b_2{\rm i}]^T$. 
    (c) Relative error $|u_{\psi}-u_t|/\mathrm{Avg}(|u_t|)$ (pink shade), with the average relative error $\varepsilon(u_{\psi},u_t)$ (black solid line).}
    \label{fig:sin_qis}
\end{figure}

We first consider a 1D target velocity field $u_t(x)=\sin{x}$. 
Note that 1D vectors will not be denoted in bold in subsequent notation. 
This simple velocity field admits an analytic wave-function solution
\begin{equation}
    \psi_1=\sin\left(\dfrac{x}{2}-\dfrac{\pi}{4}\right)e^{\mathrm{i}x},~~\psi_2=-\cos\left(\dfrac{x}{2}-\dfrac{\pi}{4}\right)e^{-\mathrm{i}x} 
\end{equation}
to \eqref{eq:psi_to_u} for validating our state preparation method. 
In Fig.~\ref{fig:sin_qis}, the optimization method can construct continuous wave functions, and approximate the target velocity field with a small relative error of $3.83\%$. 

\begin{figure}
    \subfigure{
        \begin{minipage}{0.31\linewidth}
            \centering
            \includegraphics[width=\linewidth]{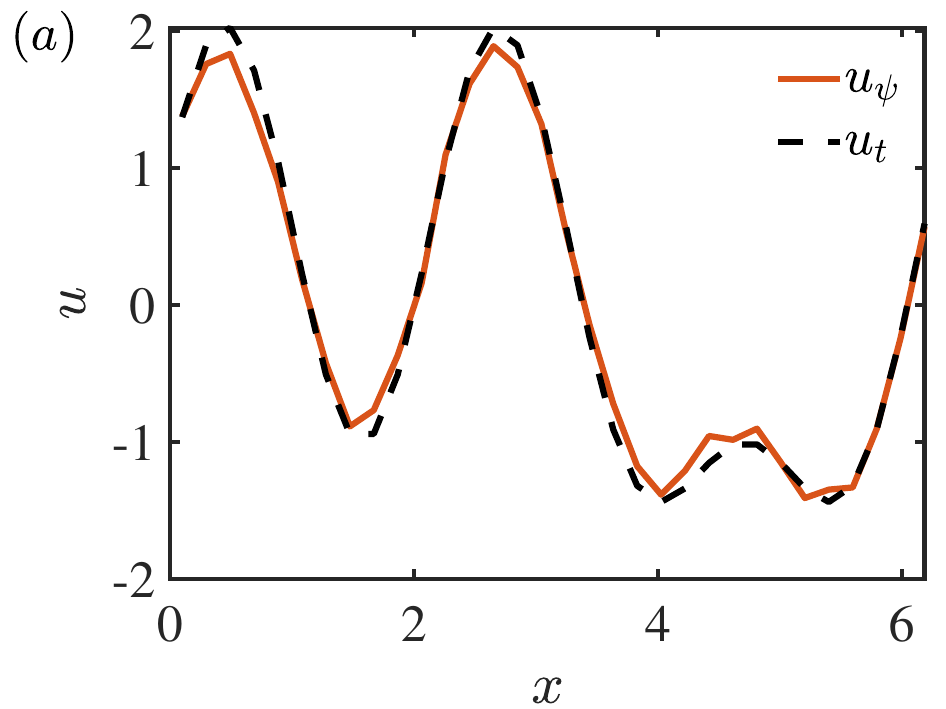}
        \end{minipage}
        \label{fig:scs_qis_u}
    }
    \subfigure{
        \begin{minipage}{0.31\linewidth}
            \centering
            \includegraphics[width=\linewidth]{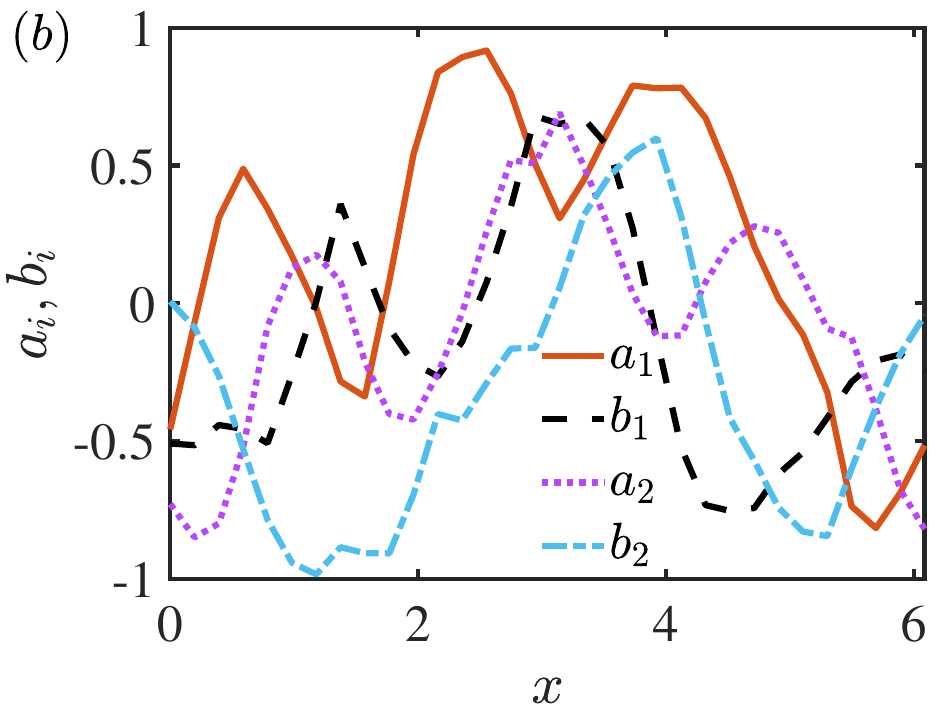}
        \end{minipage}
        \label{fig:scs_qis_abcd}
    }
    \subfigure{
        \begin{minipage}{0.31\linewidth}
            \centering
            \includegraphics[width=\linewidth]{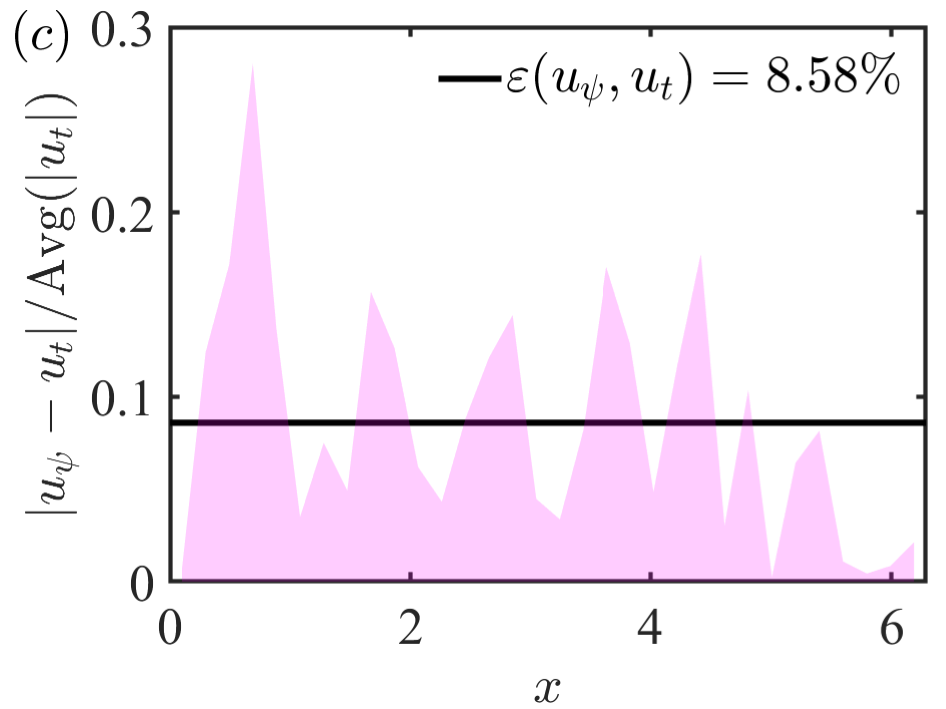}
        \end{minipage}
        \label{fig:scs_qis_dev}
    }
    \caption{Results for generating the SCWF for $u_t=\sin x+\cos 2x+\sin 3x$. 
    (a) Comparison between the target velocity $u_t$ (black dashed line) and the quantum optimization result $u_{\psi}$ (red solid line). 
    (b) Four real components of the SCWF $\bm{\psi}=[a_1+b_1{\rm i},a_2+b_2{\rm i}]^T$. 
    (c) Relative error $|u_{\psi}-u_t|/\mathrm{Avg}(|u_t|)$ (pink shade), with the average relative error $\varepsilon(u_{\psi},u_t)$ (black solid line). }
    \label{fig:scs_qis}
\end{figure}

For a more intricate target velocity $u_t(x)=\sin{x}+\cos{2x}+\sin{3x}$ with multiple Fourier modes, the results in Fig.~\ref{fig:scs_qis} shows that our algorithm adeptly generates continuous wave functions, leveraging them to approximate the target velocity with $\varepsilon = 8.58\%$.

\subsection{2D velocity fields}
We construct the quantum circuit for a 2D velocity field $\bm{u}_t=[\cos x\sin y, \sin x\cos y]$.
Note that the velocity components are coupled and exhibit diverse flow structures, such as sources, sinks, and saddle points.

\begin{figure}
    \setlength{\abovecaptionskip}{-0.5cm}
    \subfigure{
        \begin{minipage}{0.3\linewidth}
            \centering
            \includegraphics[width=\linewidth]{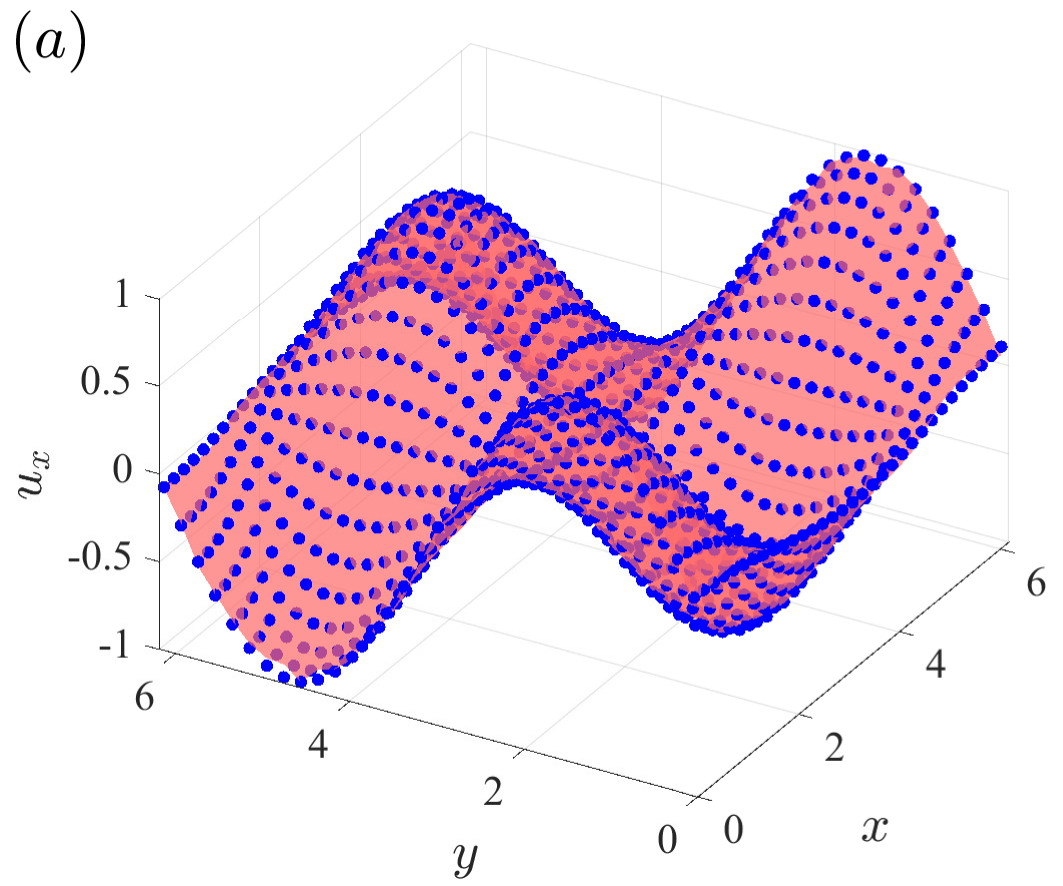}
            \label{fig:sccs_ux}
        \end{minipage}
    }
    \subfigure{
        \begin{minipage}{0.65\linewidth}
            \centering
            \includegraphics[width=\linewidth]{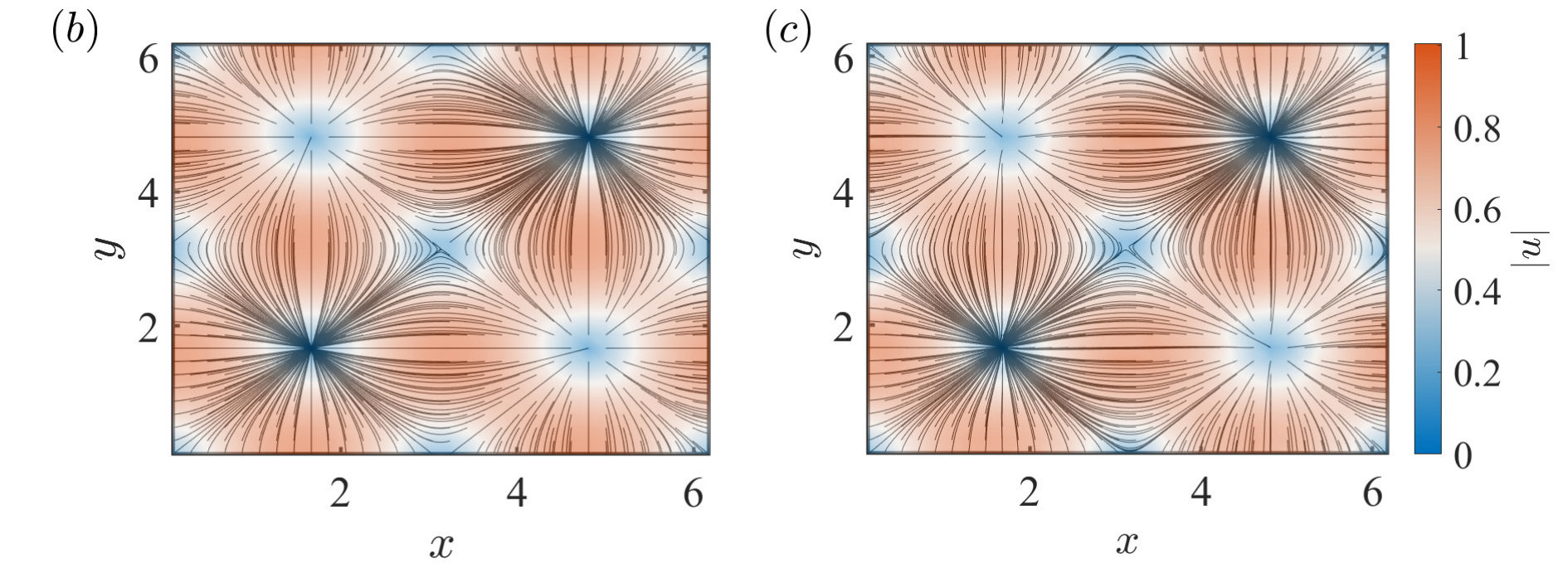}
            \label{fig:sccs_uy}
        \end{minipage}
    }
    \caption{Results for generating the SCWF for $\bm{u}_t=[\cos x\sin y, \sin x\cos y]$. 
    (a) Comparison between the target velocity $u_t$ (pink surface) and the quantum optimization result $\bm{u}_{\psi}$ (blue dots).
    Streamlines of (b) $\bm{u}_t$ and (c) $\bm{u}_{\psi}$, along with the contours of $|\bm{u}_t|$ and $|\bm{u}_{\psi}|$, respectively.}
    \label{fig:sccs_u}
\end{figure}

Figure~\ref{fig:sccs_u}(a) shows an excellent agreement between the quantum optimization result and the target field. 
The low relative error of $4.50\%$ (see Tab.~\ref{table:config}) demonstrates the capability of our algorithm in preparing quantum states for the velocity fields based on the SCWF with dimensional coupling. 
Furthermore, we visualize the streamlines of both $\bm{u}_t$ and $\bm{u}_{\psi}$ in Figs.~\ref{fig:sccs_u}(b) and \ref{fig:sccs_u}(c). 
Our algorithm successfully reproduces the intricate flow patterns induced by multiple sources, sinks, and saddle points. 

\section{Conclusions}\label{sec:conclusion}

We develop a quantum state preparation method for constructing the SCWF corresponding to a given vector field, which can represent a velocity field in fluid dynamics (or a given multidimensional vector field in other physics).  
Leveraging the pointwise normalization constraint in the SCWF, we design a quantum circuit characterized by Hadamard transformations and parameterized controlled rotation gates. Employing the VQA, we iteratively optimize the circuit parameters to facilitate the transformation of the target velocity field into the SCWF, thereby enabling subsequent quantum simulation of fluid dynamics. Through 1D and 2D case studies, we validate the accuracy and robustness of our method, highlighting its effectiveness in handling multiscale and multidimensional velocity fields. In particular, our method is able to capture essential flow topology, including sources, sinks, and saddle points, which characterize complex dynamical systems. 

On the other hand, the present method has several limitations. 
First, it has not considered the implementation on quantum hardware. 
Due to the challenges posed by quantum errors arising from decoherence and noise, implementing this algorithm on quantum hardware can be subject to computational inaccuracies. 
Second, our method encounters the exponential complexity growth with the number of qubits in circuit parameter training. Consequently, applications of this method to more complex systems, e.g., 3D flows, have expansive computational costs.
Third, the utility of the prepared quantum states has not been adequately validated in quantum simulation for flow evolution.

These limitations will be addressed to realize QCFD in future work. 
By improving the loss function for training circuit parameters and using the acceleration capabilities of quantum computing, preparation algorithms should be developed for large-scale 3D flows. Subsequently, the constructed initial state will be applied to further research on evolutionary algorithms and the development of effective quantum measurement techniques. 
The ultimate goal is to advance the SCWF-based QCFD methods \cite{Meng2023,Meng2024simulating}. 
Furthermore, the construction of SCWFs can be employed to encode quantum states for vector fields in diverse continuous media, such as magnetic fluids and elastic materials. 

\begin{acknowledgments}    
The authors thank C. Song for helpful comments.
This work has been supported in part by the National Natural Science Foundation of China (Nos.~11925201, 12302294, and 11988102), the National Key R\&D Program of China (Grant No.~2023YFB4502600), and the New Cornerstone Science Foundation through the XPLORER Prize.
\end{acknowledgments}

\appendix
\section{Sensitivity analysis of VQA parameters}\label{sec:parameters}
We investigate the influence of several critical parameters on the optimization performance with VQA, including the vortex characterization parameter $\hbar$~\cite{Chern2017}, the number $N_g$ of controlled rotation gates, and the number $N$ of grid points in each dimension. In particular, using the example shown in Fig.~\ref{fig:sin_qis_u}, we tested various parameter values to assess the algorithm's robustness and quantify ranges of the critical parameters. 
We adjust the parameter values individually to study their influence on quantum state preparation, generally adhering to those listed in Tab.~\ref{table:config}. Note that when testing different $N$ values, we chose $N_g=3$ instead of $N_g=2$ in Tab.~\ref{table:config} to ensure the quantum circuit's ability to accommodate multi-scale structures with increasing grid size. 

\begin{figure}
    \setlength{\abovecaptionskip}{-0.5cm}
    \subfigure{
        \begin{minipage}{0.45\linewidth}
            \centering
            \includegraphics[width=\linewidth]{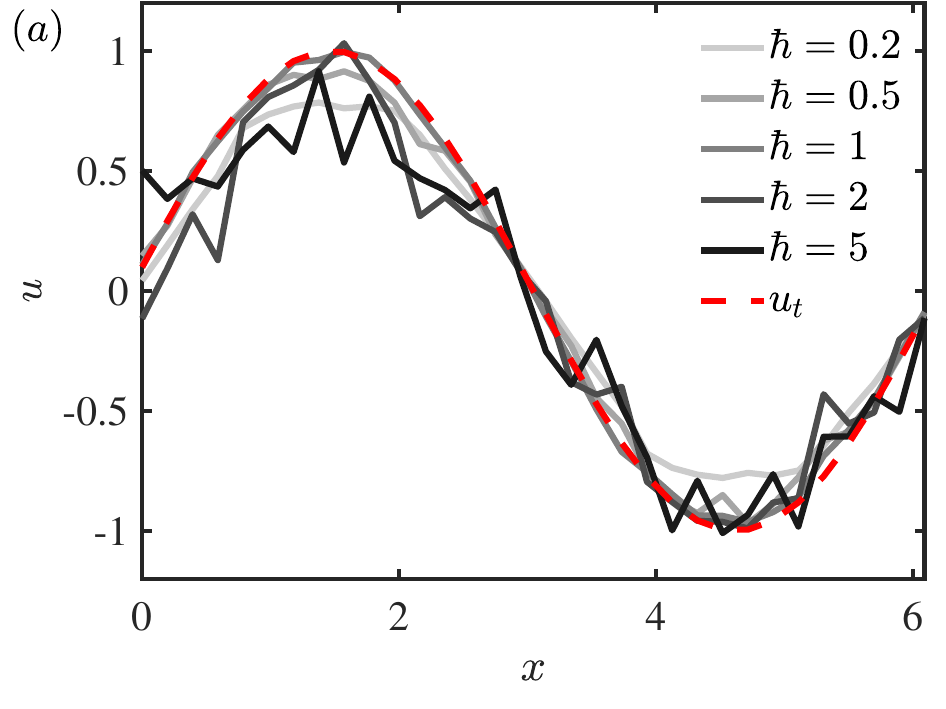}
            \label{fig:hbar_sin_u}
        \end{minipage}
    }
    \subfigure{
        \begin{minipage}{0.45\linewidth}
            \centering
            \includegraphics[width=\linewidth]{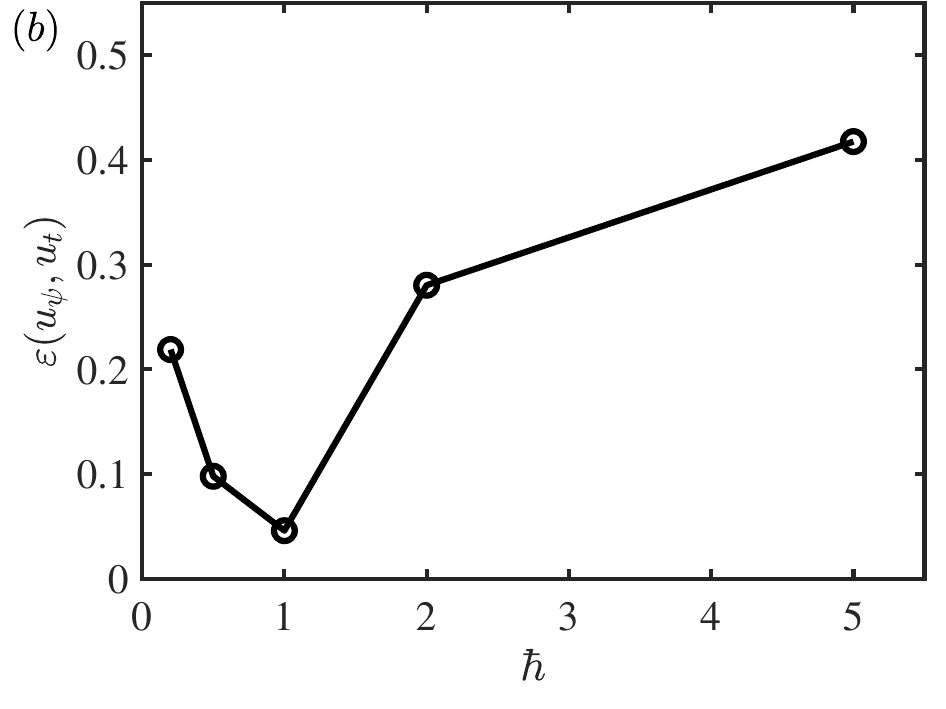}
            \label{fig:hbar_sin_dev}
        \end{minipage}
    }
    \caption{Results for generating the SCWF for $u_t=\sin x$ with $\hbar$ varying from $0.2$ to $5$. (a) Comparison between the target velocity $u_t$ (red dashed line) and the quantum optimization results $u_{\psi}$ (solid lines). (b) Relative error $\varepsilon(u_{\psi}, u_t)$ as a function of $\hbar$.} 
    \label{fig:hbar_sin}
\end{figure}

First, we conduct tests using various values of $\hbar$, including 0.2, 0.5, 1, 2, and 5. Figure \ref{fig:hbar_sin} illustrates that when $\hbar$ is within the range of approximately 0.5 to 1, the generated wave function exhibits relatively small errors. This finding is consistent with previous numerical experiments conducted on wave functions~\cite{Yang2021clebsch,Xiong2022Clebsch,Chern2016,Chern2017}. We remark that even with substantial alterations in the value of $\hbar$, such as setting $\hbar = 5$, which is 25 times its minimum value, our construction methodology can still capture the large-scale velocity profile. However, larger errors may arise, particularly near the peak velocity.

\begin{figure}
    \setlength{\abovecaptionskip}{-0.5cm}
    \subfigure{
        \begin{minipage}{0.45\linewidth}
            \centering
            \includegraphics[width=\linewidth]{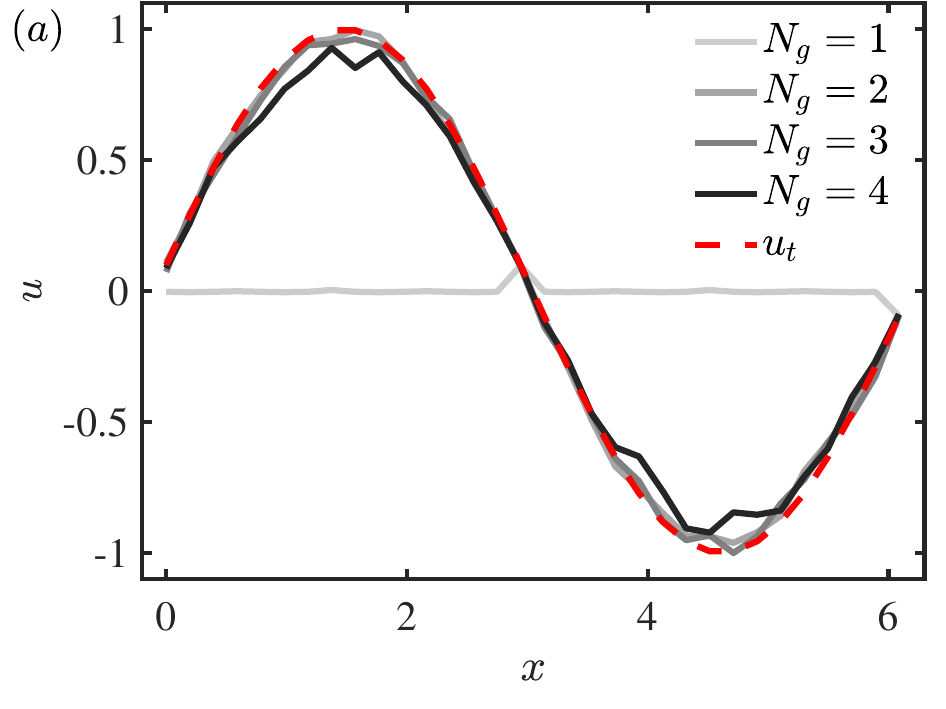}
            \label{fig:ng_sin_u}
        \end{minipage}
    }
    \subfigure{
        \begin{minipage}{0.45\linewidth}
            \centering
            \includegraphics[width=\linewidth]{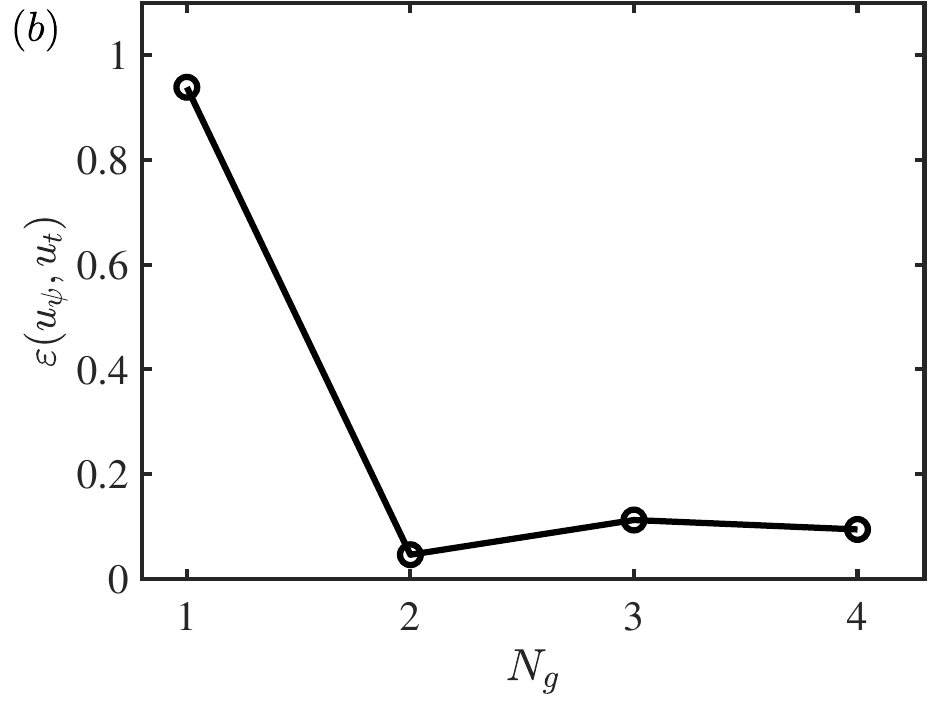}
            \label{fig:ng_sin_dev}
        \end{minipage}
    }
    \caption{Results for generating the SCWF for $u_t=\sin x$ with $N_g$ varying from $1$ to $4$. (a) Comparison between the target velocity $u_t$ (red dashed line) and the quantum optimization results $u_{\psi}$ (solid lines). (b) Relative error $\varepsilon(u_{\psi}, u_t)$ as a function of $N_g$.}
    \label{fig:ng_sin}
\end{figure}

Second, we conduct tests on $N_g$, which determines the depth of the quantum circuit, i.e., the number of layers of quantum gates. In Fig.~\ref{fig:ng_sin}, when $N_g$ is small, such as $N_g=1$, the circuit's expressiveness is limited due to the small number of adjustable parameters, resulting in large errors in the output wave function. 
With a relatively large $N_g$, such as $N_g=4$, the circuit contains sufficient adjustable parameters to approximate the target wave function. 
However, due to the circuit depth and the abundance of parameters, VQA training may encounter challenges like the barren plateau~\cite{Mcclean2018, Martin2023, Liu2022}, which hinders the convergence of parameters and leads to notable errors in the output wave function.
Hence, we suggest a moderate number of quantum circuit layers, such as $N_g=2$ and $3$.

\begin{figure}
    \setlength{\abovecaptionskip}{-0.5cm}
    \subfigure{
        \begin{minipage}{0.45\linewidth}
            \centering
            \includegraphics[width=\linewidth]{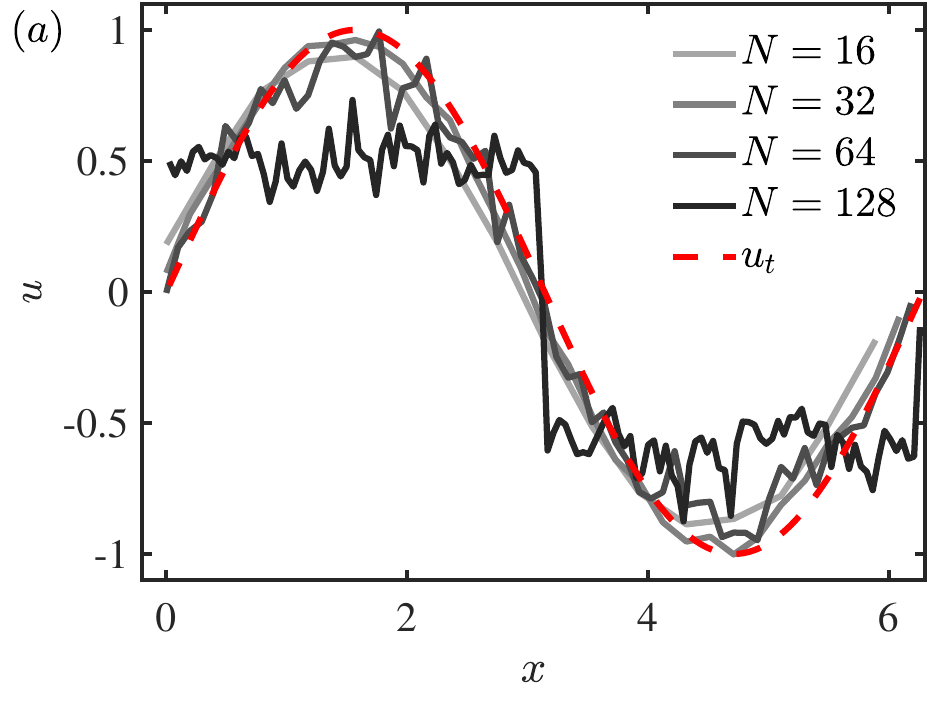}
            \label{fig:n_sin_u}
        \end{minipage}
    }
    \subfigure{
        \begin{minipage}{0.45\linewidth}
            \centering
            \includegraphics[width=\linewidth]{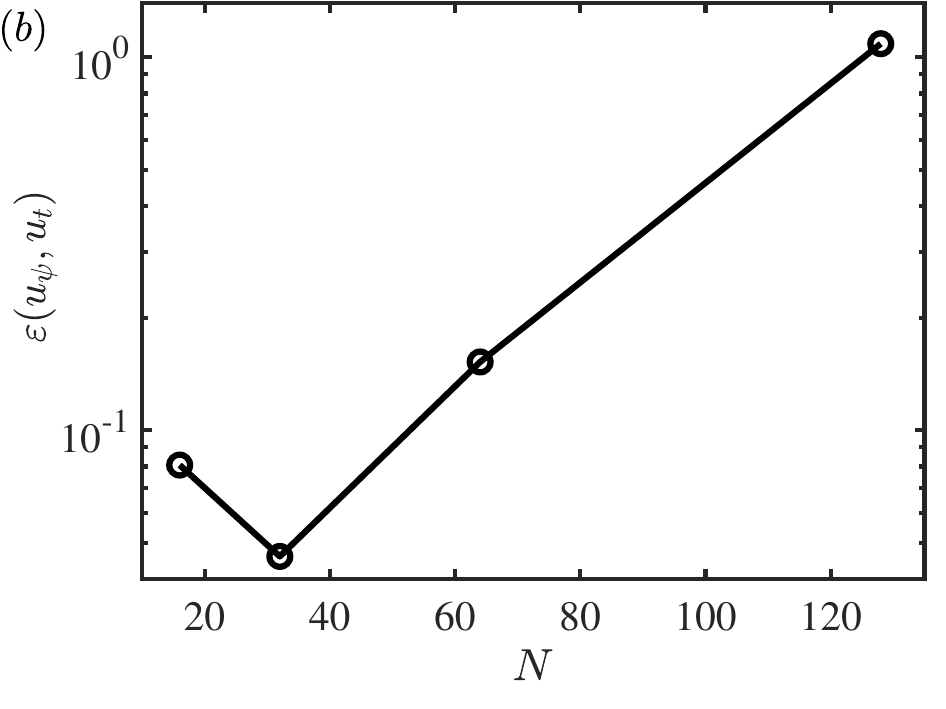}
            \label{fig:n_sin_dev}
        \end{minipage}
    }
    \caption{Results for generating the SCWF for $u_t=\sin x$ with $N$ varying from $16$ to $128$. (a) Comparison between the target velocity $u_t$ (red dashed line) and the quantum optimization results $u_{\psi}$ (solid lines). (b) Relative error $\varepsilon(u_{\psi}, u_t)$ as a function of $N$.}
    \label{fig:n_sin}
\end{figure}

Third, we conduct tests on the number of grid points. In Fig.~\ref{fig:n_sin_u}, the quantum circuit's output wave function is smoothed with decreasing $N$, due to the numerical dissipation at low grid resolutions.
Conversely, enlarging the grid size introduces more oscillations, arising from the incorporation of spurious multi-scale information by VQA. Therefore, it is essential to adjust the grid size to match the characteristic scale of the target velocity field in the SCWF-based quantum state preparation.

\begin{figure}
    \setlength{\abovecaptionskip}{-0.5cm}
    \subfigure{
        \begin{minipage}{0.45\linewidth}
            \centering
            \includegraphics[width=\linewidth]{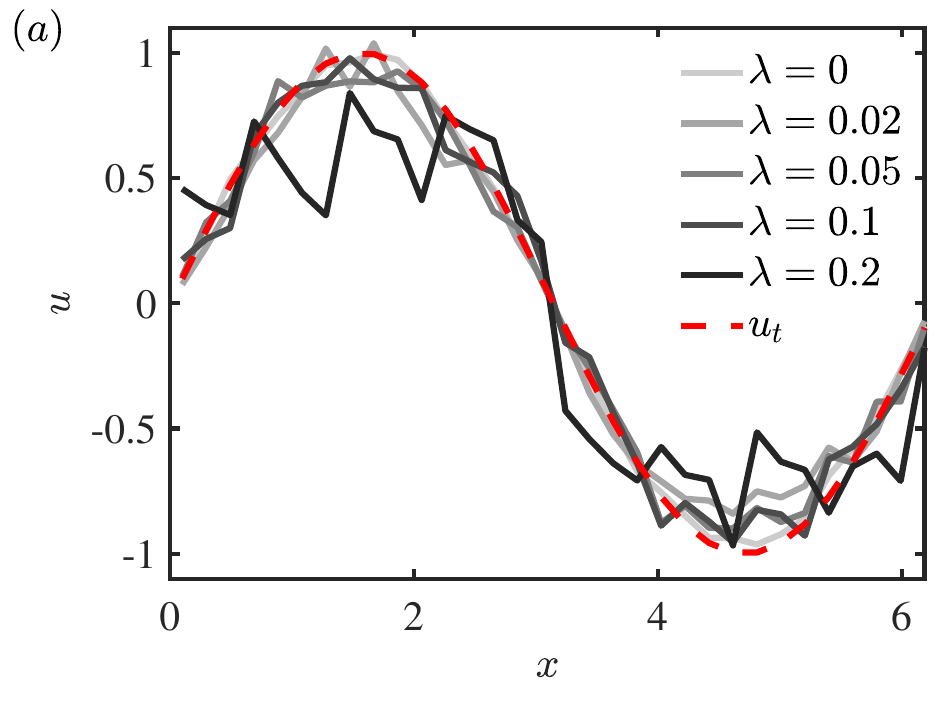}
            \label{fig:noise_sin_u}
        \end{minipage}
    }
    \subfigure{
        \begin{minipage}{0.45\linewidth}
            \centering
            \includegraphics[width=\linewidth]{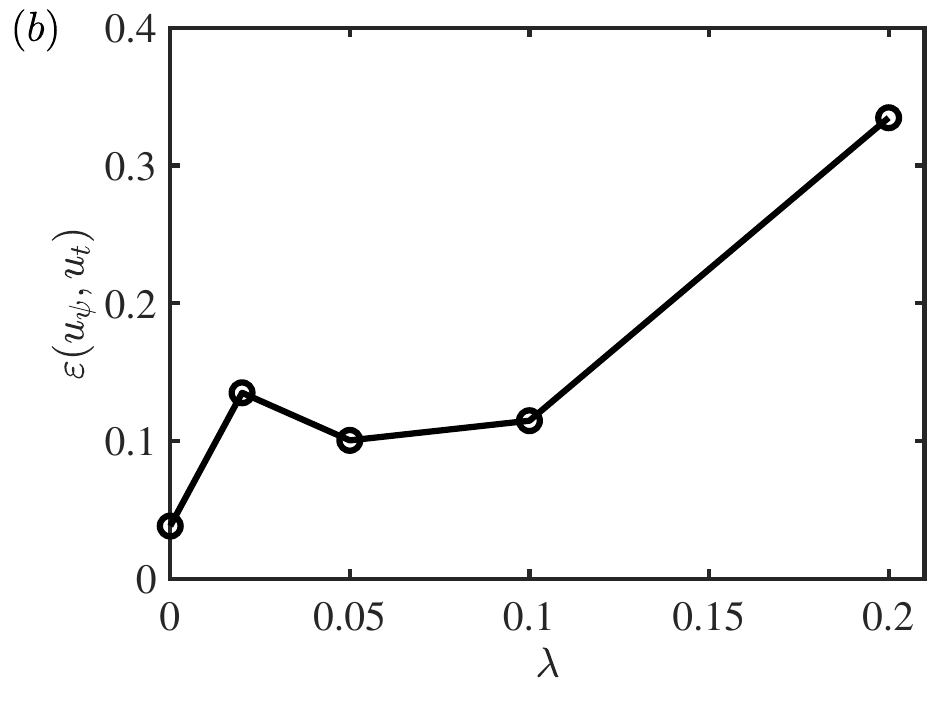}
            \label{fig:noise_sin_dev}
        \end{minipage}
    }
    \caption{Results for generating the SCWF for $u_n(x) = \sin x+\lambda\xi(-1,1)$, with varying values of $\lambda$: $0.02,0.05,0.1,0.2$. (a) Comparison between the target velocity $u_t=\sin x$ (red dashed line) and the quantum optimization results $u_{\psi}$ (solid lines). (b) Relative error $\varepsilon(u_{\psi}, u_t)$ as a function of $\lambda$.}
    \label{fig:noise_sin}
\end{figure}

Finally, we investigate the effect of noises, which are applied to the target velocity as 
\begin{equation}\label{eq:noise_target}
    u_n(x)=\sin x+\lambda \xi(-1,1).
\end{equation}
Here, $\xi(-1,1)$ is a random variable uniformly distributed on $[-1,1]$, and $\lambda$ controls the amplitude of $\xi$. 
In general, the mean error $\varepsilon(u_\psi,u_t)$ grows with $\lambda$ in \figref{fig:noise_sin_dev}, where the error is evaluated between the VQA result $u_\psi$ and the undisturbed $u_t(x)=\sin x$.

\bibliographystyle{unsrt}
\bibliography{refs}

\begin{thebibliography}{10}

\bibitem{Steane1998}
A.~Steane.
\newblock {Quantum computing}.
\newblock {\em Rep. Prog. Phys.}, 61:117, 1998.

\bibitem{Nielsen2010}
M.~A. Nielsen and I.~L. Chuang.
\newblock {\em {Quantum Computation and Quantum Information}}.
\newblock Cambridge University Press, 2010.

\bibitem{Horowitz2019}
M.~Horowitz and E.~Grumbling.
\newblock {\em {Quantum computing: progress and prospects}}.
\newblock National Academies Press, 2019.

\bibitem{Cao2019}
Y.~Cao, J.~Romero, J.~P. Olson, M.~Degroote, P.~D. Johnson, M.~Kieferov{\'a}, I.~D. Kivlichan, T.~Menke, B.~Peropadre, N.~P.~D. Sawaya, et~al.
\newblock {Quantum chemistry in the age of quantum computing}.
\newblock {\em Chem. Rev.}, 119:1085, 2019.

\bibitem{Wittek2014}
P.~Wittek.
\newblock {\em {Quantum machine learning: what quantum computing means to data mining}}.
\newblock Academic Press, 2014.

\bibitem{Mavroeidis2018}
V.~Mavroeidis, K.~Vishi, M.~D. Zych, and A.~J{\o}sang.
\newblock The impact of quantum computing on present cryptography.
\newblock {\em arXiv preprint arXiv:1804.00200}, 2018.

\bibitem{Schuld2015}
M.~Schuld, I.~Sinayskiy, and F.~Petruccione.
\newblock {An introduction to quantum machine learning}.
\newblock {\em Contemp. Phys.}, 56:172--185, 2015.

\bibitem{Jin2023}
S.~Jin, N.~Liu, and Y.~Yu.
\newblock {Quantum simulation of partial differential equations: Applications and detailed analysis}.
\newblock {\em Phys. Rev. A}, 108:032603, 2023.

\bibitem{Gaitan2020}
F.~Gaitan.
\newblock {Finding flows of a Navier-Stokes fluid through quantum computing}.
\newblock {\em NPJ Quantum Inf.}, 6:61, 2020.

\bibitem{Shende2004}
V.~V. Shende, I.~L. Markov, and S.~S. Bullock.
\newblock Minimal universal two-qubit controlled-not-based circuits.
\newblock {\em Phys. Rev. A}, 69:062321, 2004.

\bibitem{Shende2006}
V.~V. Shende, S.~S. Bullock, and I.~L. Markov.
\newblock Synthesis of quantum-logic circuits.
\newblock {\em IEEE Transactions on Computer-Aided Design of Integrated Circuits and Systems}, 25:1000--1010, 2006.

\bibitem{Drury2008}
B.~Drury and P.~Love.
\newblock Constructive quantum shannon decomposition from cartan involutions.
\newblock {\em J. Phys. A: Math. Theor.}, 41:395305, 2008.

\bibitem{Khammassi2021}
N.~Khammassi, I.~Ashraf, J.~V. Someren, R.~Nane, A.~M. Krol, M.~A. Rol, L.~Lao, K.~Bertels, and C.~G. Almudever.
\newblock Openql: A portable quantum programming framework for quantum accelerators.
\newblock {\em ACM J. Emerg. Tech. Com.}, 18:1--24, 2021.

\bibitem{Preskill2018}
J.~Preskill.
\newblock {Quantum computing in the NISQ era and beyond}.
\newblock {\em Quantum}, 2:79, 2018.

\bibitem{Bharti2022}
K.~Bharti, A.~Cervera-Lierta, T.~H. Kyaw, T.~Haug, S.~Alperin-Lea, A.~Anand, M.~Degroote, H.~Heimonen, J.~S. Kottmann, T.~Menke, et~al.
\newblock {Noisy intermediate-scale quantum algorithms}.
\newblock {\em Rev. Mod. Phys.}, 94:015004, 2022.

\bibitem{Khatri2019}
S.~Khatri, R.~LaRose, A.~Poremba, L.~Cincio, A.~T. Sornborger, and P.~J. Coles.
\newblock Quantum-assisted quantum compiling.
\newblock {\em Quantum}, 3:140, 2019.

\bibitem{Batchelor1967}
G.~K. Batchelor.
\newblock {\em {An Introduction to Fluid Dynamics}}.
\newblock Cambridge University Press, 1967.

\bibitem{nair2016elements}
V.~P. Nair.
\newblock Elements of geometric quantization and applications to fields and fluids.
\newblock {\em arXiv preprint arXiv:1606.06407}, 2016.

\bibitem{Ray2022}
N.~Ray, T.~Banerjee, B.~Nadiga, and S.~Karra.
\newblock {On the Viability of Quantum Annealers to Solve Fluid Flows}.
\newblock {\em Front Mech. Eng.}, 8:906696, 2022.

\bibitem{Pfeffer2022}
P.~Pfeffer, F.~Heyder, and J.~Schumacher.
\newblock {Hybrid quantum-classical reservoir computing of thermal convection flow}.
\newblock {\em Phys. Rev. Res.}, 4:033176, 2022.

\bibitem{Harrow2009}
A.~W. Harrow, A.~Hassidim, and S.~Lloyd.
\newblock {Quantum algorithm for linear systems of equations}.
\newblock {\em Phys. Rev. Lett.}, 103:150502, 2009.

\bibitem{Meng2023}
Z.~Y. Meng and Y.~Yang.
\newblock {Quantum computing of fluid dynamics using the hydrodynamic Schr\"{o}dinger equation}.
\newblock {\em Phys. Rev. Research}, 5:033182, 2023.

\bibitem{Meng2024quantum}
Z.~Meng and Y.~Yang.
\newblock Quantum spin representation for the navier-stokes equation.
\newblock {\em arXiv preprint arXiv:2403.00596}, 2024.

\bibitem{Chern2016}
A.~Chern, F.~Kn\"{o}ppel, U.~Pinkall, P.~Schr\"{o}der, and S.~Wei\ss{}mann.
\newblock {Schr\"{o}dinger's smoke}.
\newblock {\em ACM Trans. Graph.}, 35:77, 2016.

\bibitem{Chern2017}
A.~Chern, F.~Kn\"{o}ppel, U.~Pinkall, and P.~Schr\"{o}der.
\newblock {Inside fluids: Clebsch maps for visualization and processing}.
\newblock {\em ACM Trans. Graph.}, 36:142, 2017.

\bibitem{Yang2021clebsch}
S.~Yang, S.~Xiong, Y.~Zhang, F.~Feng, J.~Liu, and B.~Zhu.
\newblock {Clebsch gauge fluid}.
\newblock {\em ACM Trans. Graph.}, 40:99, 2021.

\bibitem{Xiong2022Clebsch}
S.~Xiong, Z.~Wang, M.~Wang, and B.~Zhu.
\newblock {A Clebsch method for free-surface vortical flow simulation}.
\newblock {\em ACM Trans. Graph.}, 41:4, 2022.

\bibitem{Tao2021}
R.~Tao, H.~Ren, Y.~Tong, and S.~Xiong.
\newblock {Construction and evolution of knotted vortex tubes in incompressible Schr\"{o}dinger flow}.
\newblock {\em Phys. Fluids}, 33:077112, 2021.

\bibitem{Benedetti2019}
M.~Benedetti, E.~Lloyd, S.~Sack, and M.~Fiorentini.
\newblock {Parameterized quantum circuits as machine learning models}.
\newblock {\em Quantum Sci. Technol.}, 4:043001, 2019.

\bibitem{Jaksch2023}
D.~Jaksch, P.~Givi, A.~J. Daley, and T.~Rung.
\newblock {Variational quantum algorithms for computational fluid dynamics}.
\newblock {\em AIAA J.}, 61:1885--1894, 2023.

\bibitem{Chern2017thesis}
A.~Chern.
\newblock {\em {Fluid dynamics with incompressible Schr\"{o}dinger flow}}.
\newblock PhD thesis, California Institute of Technology, 2017.

\bibitem{Hopf1931}
H.~Hopf.
\newblock {{\"U}ber die Abbildungen der dreidimensionalen Sph{\"a}re auf die Kugelfl{\"a}che}.
\newblock {\em Math. Ann.}, 104:637--665, 1931.

\bibitem{Bloch1946}
F.~Bloch.
\newblock {Nuclear induction}.
\newblock {\em Phys. Rev.}, 70:460, 1946.

\bibitem{Yang2010b}
Y.~Yang and D.~I. Pullin.
\newblock {On Lagrangian and vortex-surface fields for flows with Taylor--Green and Kida--Pelz initial conditions}.
\newblock {\em J. Fluid Mech.}, 661:446--481, 2010.

\bibitem{Yang2011b}
Y.~Yang and D.~I. Pullin.
\newblock {Evolution of vortex-surface fields in viscous Taylor--Green and Kida--Pelz flows}.
\newblock {\em J. Fluid Mech.}, 685:146--164, 2011.

\bibitem{Loshchilov2019}
L.~Ilya and H.~Frank.
\newblock Decoupled weight decay regularization.
\newblock {\em arXiv preprint arXiv:1711.05101}, 2019.

\bibitem{Lockwood2022}
O.~Lockwood.
\newblock An empirical review of optimization techniques for quantum variational circuits.
\newblock {\em arXiv preprint arXiv:2202.01389}, 2022.

\bibitem{Mcclean2018}
J.~R. McClean, S.~Boixo, V.~N. Smelyanskiy, R.~Babbush, and H.~Neven.
\newblock {Barren plateaus in quantum neural network training landscapes}.
\newblock {\em Nat. Commun.}, 9:4812, 2018.

\bibitem{Martin2023}
E.~C. Mart{\'\i}n, K.~Plekhanov, and M.~Lubasch.
\newblock {Barren plateaus in quantum tensor network optimization}.
\newblock {\em Quantum}, 7:974, 2023.

\bibitem{Liu2022}
Z.~Liu, L.~W. Yu, L.~M. Duan, and D.~L. Deng.
\newblock {Presence and absence of barren plateaus in tensor-network based machine learning}.
\newblock {\em Phys. Rev. Lett.}, 129:270501, 2022.

\bibitem{Meng2024simulating}
Z.~Meng, J.~Zhong, S.~Xu, K.~Wang, J.~Chen, F.~Jin, X.~Zhu, Y.~Gao, Y.~Wu, C.~Zhang, N.~Wang, Y.~Zou, A.~Zhang, Z.~Cui, F.~Shen, Z.~Bao, Z.~Zhu, Z.~Tan, T.~Li, P.~Zhang, S.~Xiong, H.~Li, Q.~Guo, Z.~Wang, C.~Song, H.~Wang, and Y.~Yang.
\newblock Simulating unsteady fluid flows on a superconducting quantum processor.
\newblock {\em arXiv preprint arXiv:2404.15878}, 2024.

\end{thebibliography}

\end{document}